\documentclass[floatfix,twocolumn,aps,prl,draft,showpacs]{revtex4}
\usepackage{graphicx}
\usepackage{bm}
\begin{document}
\title{Turbulence and coarsening in active and passive binary mixtures}
\author{S. Berti$^1$, G. Boffetta$^1$, M. Cencini$^{2,4}$, A.
Vulpiani$^{3,4}$} \affiliation{ $^1$ Dipartimento di Fisica Generale
and INFN Universit\`a di Torino, Via Pietro Giuria 1, I-10125 Torino,
Italy.\\ $^{2}$ ISC-CNR, Via dei Taurini 19, I-00185 Roma, Italy.\\
$^{3}$ Dipartimento di Fisica and INFN Sezione Universit\`a di Roma
``La Sapienza'', P.zzle Aldo Moro 2, I-00185 Rome, Italy.\\
$^4$INFM-SMC P.zzle Aldo Moro 2, I-00185 Rome, Italy.}  \date{\today}
\begin{abstract}
Phase separation between two fluids in two-dimensions is investigated
by means of Direct Numerical Simulations (DNS) of coupled
Navier-Stokes and Cahn-Hilliard equations.  We study the phase
ordering process in the presence of an external stirring acting on the
velocity field.  For both active and passive mixtures we find that,
for a sufficiently strong stirring, coarsening is arrested in a
stationary dynamical state characterized by a continuous rupture and
formation of finite domains.  Coarsening arrest is shown to be
independent of the chaotic or regular nature of the flow.
\end{abstract}
 \maketitle 

\begin{figure}[t!]  
\includegraphics[draft=false, scale=0.37, clip=true]{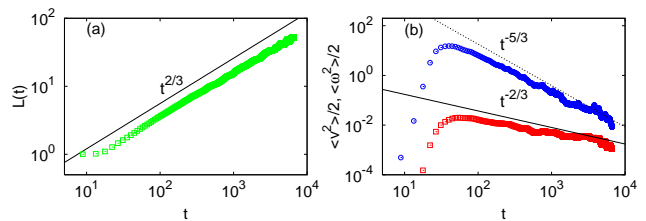}
\caption{(color online) (a) $L(t)$ vs $t$ obtained by DNS of
  (\protect\ref{eq:ch}) and (\protect\ref{eq:ns}) with $\xi=0.015$,
  $\nu=10^{-3}$, without external forcing.  (b) Kinetic energy ${\cal
  K}=\langle v^2\rangle/2$ (bottom) and enstrophy
  $\Omega=\langle \omega^2 \rangle/2$ (top) vs $t$ in the
  same run.\label{fig:unstirred}}
\end{figure}

\begin{figure}[b!]
\includegraphics[draft=false, scale=0.27, clip=true]{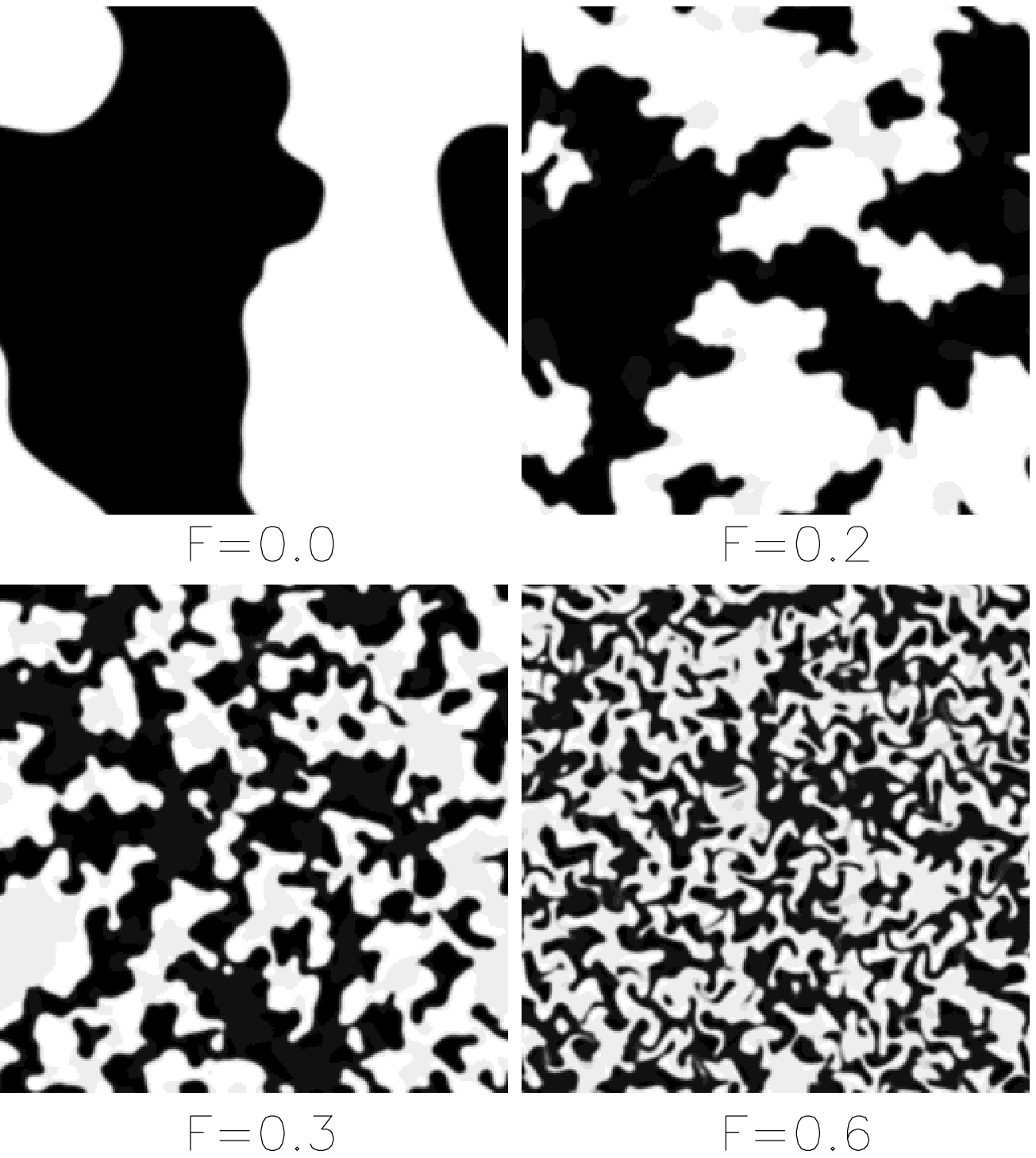} 
\hspace{1truecm}
\includegraphics[draft=false, scale=0.27, clip=true]{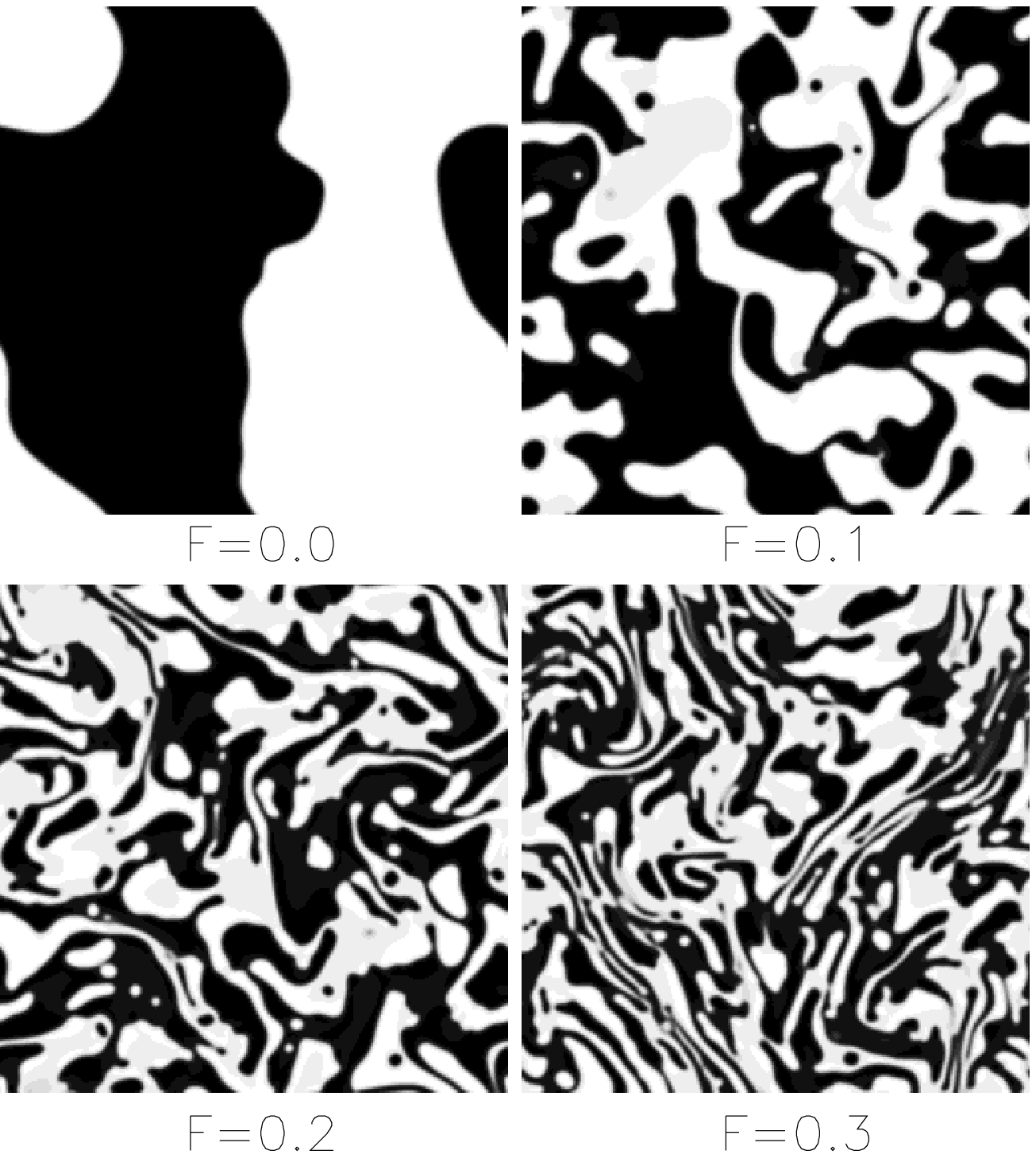} 
\caption{Snapshots of $\theta$ at time $t=4000$ at varying the forcing
  intensity $F$ with $\ell_f\!=\!26 \xi$ (left) and $\ell_f\!=\!84 \xi$
  (right). Black/white codes $\theta=\pm 1$. Other parameters are set as in
  Fig.~\protect\ref{fig:unstirred}}\label{fig:snaps_attivo}
\end{figure}

\begin{figure}[t!]
\includegraphics[draft=false, scale=0.27, clip=true]{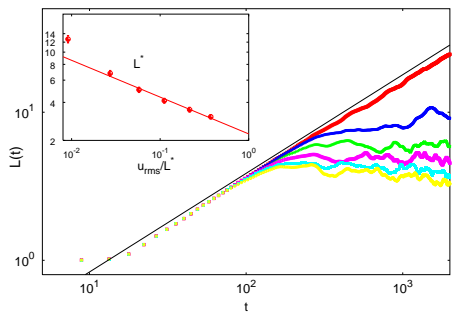} 
\caption{(color online) $L$ vs $t$ at varying $F$, from top
$F\!\!=\!\!0$ (thick curve) and $F\!\!=\!\! 0.05,
0.10,0.15,\dots,0.30$. Data refer to DNS with $\ell_f=84 \xi$ (the
case with $\ell_f=26 \xi$ is qualitatively similar). The straight line
displays the scaling $t^{2/3}$. Inset: $L^*$ vs $u_{rms}/L^*$; the
straight line has slope $-0.29$, point size is of the order of the
statistical error.}
\label{fig:lengths_attivo}
\end{figure}

When a binary fluid mixture at the critical concentration is cooled
from a high temperature to a sufficiently low temperature (below a
critical one), the original homogeneous phase becomes unstable and
spontaneously evolves into two phases separated by an interface. As
time advances an out-of-equilibrium process of phase ordering takes
place through the formation of domains of a single phase that grow
algebraically in time~\cite{F85B94}. In fluids, the presence of a
hydrodynamic velocity field makes this process more complicated than
the corresponding one in solid alloys. For instance, since Siggia's
seminal work~\cite{siggia79}, it is well known that hydrodynamics may
accelerate the domain growth (see Refs.~\cite{KCPDB01}
and~\cite{WY98-Y99} for recent development in three and two
dimensional fluids, respectively).  Phase ordering dynamics becomes
even more complex and less understood when the fluid mixture is
externally driven~\cite{O97,RN81-AN84}; beyond their theoretical
interest, phase separating binary fluids under flow embody a great
technological interest~\cite{O84} for their distinctive rheological
properties. This problem has been extensively investigated in shear
flows~\cite{CKBD99,CGL98,B01,HMM95} where coarsening becomes highly
anisotropic: the single-phase domain growth accelerates in the shear
direction while in the transversal one the growth is
arrested~\cite{B01,HMM95} or strongly slowed down~\cite{CGL98}. Less
clear is the case in which the mixture is stirred by a turbulent
flow~\cite{O84,PEMG84,CGM87-TGSO89,MSPG89}. Here, phase separation may
be completely suppressed~\cite{CGM87-TGSO89}, or a dynamical steady
state with domains of finite length and well defined phases may
develop~\cite{RN81-AN84,MSPG89}.  A similar phenomenology has been
experimentally observed in stirred immiscible fluids~\cite{Ottino}.
This Letter aims to clarify the nature of the non-equilibrium steady
state, characterized by the continuous rupture and formation of
domains.\\ \indent Previous investigations focused on {\it passive}
binary mixtures (when the feedback of the phase ordering on fluid
velocity is neglected) in random flows~\cite{LSS95} and in chaotic
flows \cite{BBK01} (in a Lagrangian sense, i.e. two initially very
close particles separate exponentially in time \cite{CFPV91}).  Here
we focus on the phase ordering dynamics of {\it active}
two-dimensional binary mixtures in which the fluid is driven by
chemical potential inhomogeneities~\cite{RN81-AN84}.  By means of
numerical experiments we show that coarsening arrest is a generic and
robust phenomenon, whose existence can be understood by an energy
conservation argument. Moreover, we show that in the passive limit
Lagrangian Chaos is not necessary for coarsening arrest. \\\indent In
the presence of stirring, the main question concerns the competition
between thermodynamic forces, driving the phase segregation, and fluid
motion, leading to mixing and domains break-up.  For very high flow
intensities phase separation can be completely
suppressed~\cite{PEMG84,CGM87-TGSO89} due to mixing of the components
and inhibition of interface formation. In active mixtures with very
low viscosities such a phenomenon may be self-induced by the
feedback~\cite{T94-TA98,WY98-Y99}:  the fluid responds vigorously to
local chemical potential variations and re-mixes the components. On
the other hand, stirring may lower the critical
temperature~\cite{SR85,MSPG89,O97}. However, by performing a
deeper quench, phase separation in a nontrivial statistically
stationary state may still develop~\cite{RN81-AN84}.\\\indent Being
interested in deep quenching, here we work at zero temperature, as in
Refs.~\cite{LSS95,BBK01}.  We consider a symmetric ($50\%-50\%$)
mixture of two incompressible fluids of equal density $\rho=1$ and
viscosity $\nu$. Such a bi-component system is described by a scalar
order parameter $\theta(\bm r,t)$, the local fraction of the two
fluids. The associated Landau-Ginzburg free energy
reads~\cite{F85B94}:
\begin{equation}
\Phi[\theta]\! =\!\int {\rm d}{\bm r}
\left( -{1\over 2} \theta^2 +{1\over 4} \theta^4 + {\xi^2 \over 2}
|{\bm \nabla} \theta|^2 \right)\,,
\end{equation}
being $\xi$ the equilibrium correlation length, which provides a measure of
the interface width.  The dynamics is then governed by the Cahn-Hilliard (CH)
equation
\begin{equation}
\partial_t \theta+{\bm v}\cdot{\bm \nabla} \theta = \Gamma \nabla^2
{\delta \Phi\over \delta \theta}= \Gamma \nabla^2 \mu
\label{eq:ch}
\end{equation}
where $\mu=- \theta+\theta^3-\xi^2 \nabla^2 \theta$ is the chemical
potential, and $\Gamma$ is a mobility coefficient that we assume
constant and independent of $\theta$. Hydrodynamics enters in
Eq.~(\ref{eq:ch}) through the convective term. The order parameter is
transported by the two-dimensional velocity field $\bm v$, which
evolves according to the Navier-Stokes (NS) equation
\begin{equation}
\partial_t {\bm v}+{\bm v}\cdot{\bm \nabla} {\bm v}=\nu
\nabla^2{\bm v}- {\bm \nabla}p-\theta {\bm \nabla}\mu+ {\bm f}\,,
\label{eq:ns}
\end{equation}
where $p$ is the pressure. The fluid is forced by the external
mechanical force $f$ and by local chemical potential variations
$-\theta {\bm \nabla}\mu$ (the two phases want to demix and thus force
the fluid, see also Ref.~\cite{KCPDB01} for a detailed derivation).  
This latter term can be rewritten as $-\xi^2
\nabla^2\theta {\bm \nabla}\theta$ plus a gradient term which can be
absorbed into the pressure \cite{RN81-AN84}.  Therefore
Eq.(\ref{eq:ns}) formally reduces to the 2D magnetohydrodynamics (MHD)
equation for the velocity field.
Actually phase ordering and MHD share many phenomenological properties
\cite{RN81-AN84}.  \\\indent We numerically integrate the coupled equations
(\ref{eq:ch})-(\ref{eq:ns}) by means of a standard pseudo-spectral
code implemented on a two-dimensional periodic box of size $2\pi\times
2\pi$ with $512^2$ collocation points. 
Statistical analysis of domain sizes is obtained by
considering the characteristic length, defined as
$L(t)=\langle(1-\theta^2)\rangle^{-1}$ where $\langle ... \rangle$
denotes spatial average~\cite{nota}.  The initial condition for
the order parameter is a high temperature configuration with $\theta$
set as white noise in space.  In the presented results time is
rescaled with the diffusive time $t_m=\xi^2/\Gamma$. \\\indent {\it
Unstirred case.} Starting from the initial configuration with the
fluid at rest ($\bm v=0$), after a few diffusive time scales
$t_m$, sharp interfaces appear and phase separation proceeds through
domain coarsening.  At long times, the domains length $L$ -- the only
characteristic scale of the system (provided $L\gg \xi$) -- grows in
time as a power law~\cite{F85B94}. In 2D different regimes have been
predicted and observed~\cite{WY98-Y99}: $L(t)\sim t^{1/3}$, as in
fluids at rest, for high viscosity; $L(t)\sim t^{2/3}$ for lower
viscosities.  At intermediate values of viscosity it is still unclear
whether there is only one characteristic scale~\cite{WY98-Y99}; for
$\nu \ll 1$ and low mobility $\Gamma \ll 1$ mixing may overwhelm phase
demixing~\cite{T94-TA98}.  In the following we will limit our analysis
to the turbulent, low viscous, regime where scaling exponent $2/3$ is
expected.  This exponent can be dimensionally derived by balancing the
inertial term $\bm v\cdot \bm \nabla \bm v$ with $\theta\bm\nabla
\mu$ in (\ref{eq:ns}), and assuming that $L(t)$ is the only length scale of the
system. The scaling behavior of $L(t)$ implies the
following ones for the kinetic energy and the
enstrophy~\cite{KCPDB01}: ${\cal K}=\langle v^2\rangle/2 \sim
t^{-2/3}$ and $\Omega=\langle \omega^2 \rangle/2\sim t^{-5/3}$ 
($\omega=\bm \nabla \times \bm v$ is the vorticity).
Fig.~\ref{fig:unstirred} shows that the scaling predictions are well
reproduced by our DNS.  We remark that in the absence of stirring,
phase separation is accelerated by the presence of hydrodynamics.\\
\indent{\it Stirred case.} We now consider the presence of an external
mechanical forcing acting on the velocity field. As customary in
turbulent simulations, energy is injected by means of a random, time
uncorrelated, homogeneous, and isotropic Gaussian process with amplitude
$F$ which is restricted to a few Fourier modes around $k_f$ (this
identifies the injection scale $\ell_f\sim 2\pi/k_f$).  The
$\delta$-correlation in time allows for controlling the kinetic energy
input $\epsilon_{in}\!\!=\!\!F^2 n_f$ (being $n_f$ the number of
excited Fourier modes).  Equations (\ref{eq:ch})-(\ref{eq:ns}) are
integrated starting with $\bm v=0$.  In Fig.~\ref{fig:snaps_attivo} we
show typical snapshots of the order parameter at different flow
intensities, and for two random forcing with $\ell_f\approx 26\xi$
and $\ell_f \approx 84\xi$.  As it is evident, the
stronger the stirring intensity the smaller the typical domain length.
This is confirmed by the temporal evolution of $L(t)$ at varying the
external forcing (Fig.~\ref{fig:lengths_attivo}).  After an initial
growth characterized by the $2/3$ scaling exponent, $L(t)$ stabilizes
at a value $L^*$ that decreases with the stirring intensity. Both the
kinetic energy ${\cal K}(t)$ and the enstrophy $\Omega(t)$ (not shown
here) stabilize at corresponding values.  Therefore a well defined
statistically steady state is reached.
\\\indent A closer inspection of Fig.~\ref{fig:snaps_attivo} reveals
qualitative differences in domain shapes.  When $L^*$ is larger than
the forcing scale $\ell_f$ (left) the domains are almost isotropic,
while in the case $L^{*}<\ell_f$ (right) the underlying velocity field
reveals itself through the filamental structure of the domains.
Nevertheless, coarsening process is always arrested confirming the
robustness of the phenomenon.  Stirring always selects a scale through
the competition between the thermodynamic forces and the stretching
induced by local shears that deform and break the domains.
Estimating the shear rate as $\gamma=u_{rms}/L^*$ we find that, for
the case with $L^{*}<\ell_f$, $L^*\sim \gamma^{-0.29}$ (see inset of
Fig.~\ref{fig:lengths_attivo}), in fairly good agreement with
experiments and simulations in pure shear flows \cite{HMM95}.  However
we should mention that in our settings, homogeneous and isotropic
flows, there is not a well defined rate $\gamma$ as in genuine shear
flows. The definition adopted here is a dimensional estimation of the
shear rate at the arrest scale. In the case $L^{*}\geq \ell_f$ no
clear scaling behavior is observed.\\ \indent The existence of a
stationary state can be understood in terms of conservation laws. Due
to the presence of two inviscid quadratic invariants, ${\cal K}$ and
${\Omega}$, the single fluid 2D NS equation (i.e. (\ref{eq:ns})
without the feedback term) is characterized by a double
cascade~\cite{K67}: ${\cal K}$ flows toward the large scales
($r\!>\!\ell_f$) and $\Omega$ toward the small ones
($r\!\!<\!\!\ell_f$).  By switching on the coupling term, the
following balance equation for the total energy $E={\cal K} +\Phi$
holds~\cite{nota2}
\begin{equation}
{d E\over dt}=  
\!-\nu \langle|{\bm \nabla v}|^2 \rangle 
\!-\!\Gamma \langle|{\bm \nabla}\mu|^2 \rangle \!+\epsilon_{in}\,.
\label{eq:bilancio}
\end{equation}
Now it is clear that in the unstirred case ($\epsilon_{in}\!=\!\!0$)
asymptotically an equilibrium state, corresponding to fluid at rest
$v\!=\!0$ and complete phase separation (minimum of free energy), is
reached. In this case the velocity has only a transient role,
determining the scaling of the coarsening process. On the contrary if
$\epsilon_{in}\!\neq\! 0$ a nontrivial stationary state stems from the
balance of dissipative and input terms in the
r.h.s. of~(\ref{eq:bilancio}).  \\\indent It is worth mentioning that
when the stirring intensity becomes high enough to overcome the
feedback term, the kinetic energy dissipation induced by the $|\bm
\nabla \mu|^2$ term is no more effective.  Indeed, when $\ell_f$ is
much larger than $\xi$ and $F$ is very high, the coupling term becomes
negligible and we observe the single-fluid phenomenology with an
inverse energy cascade.  \\\indent {\it Passive binary mixtures} We
now consider the case in which the coupling term in (\ref{eq:ns}) is
switched off and consequently the order parameter is passively
transported by the velocity field. This case has been already
considered in \cite{LSS95,BBK01}.  In order to obtain a statistically
stationary state, as customary we added a large scale friction term
$-\alpha \bm v$ to the Navier-Stokes equation~\cite{NOAG00}.  The
velocity field in (\ref{eq:ch}) is rescaled by a factor $\beta$; this
is a numerically convenient way to change the velocity intensity and
to study the effect of stirring on coarsening.  For $\beta=0$
Eq.~({\ref{eq:ch}) recovers the Cahn-Hilliard equation in a fluid at
rest for which $L(t)\sim t^{1/3}$.  For $\beta>0$ we observe the
following phenomenology (Fig.~\ref{fig:length_passive}a). For small
values of $\beta$ (weak stirring) we did not find a clear evidence of
coarsening arrest. This is likely due to finite size effects hiding
the phenomenon, i.e. $L^*$ becomes comparable with, or even larger
than, the box size. For $\beta$ large enough (strong stirring), the
existence of an arrest scale $L^*$ (that decreases with $\beta$) is
well evident. \\
\begin{figure}[t!]
\includegraphics[draft=false, scale=0.32,clip=true]{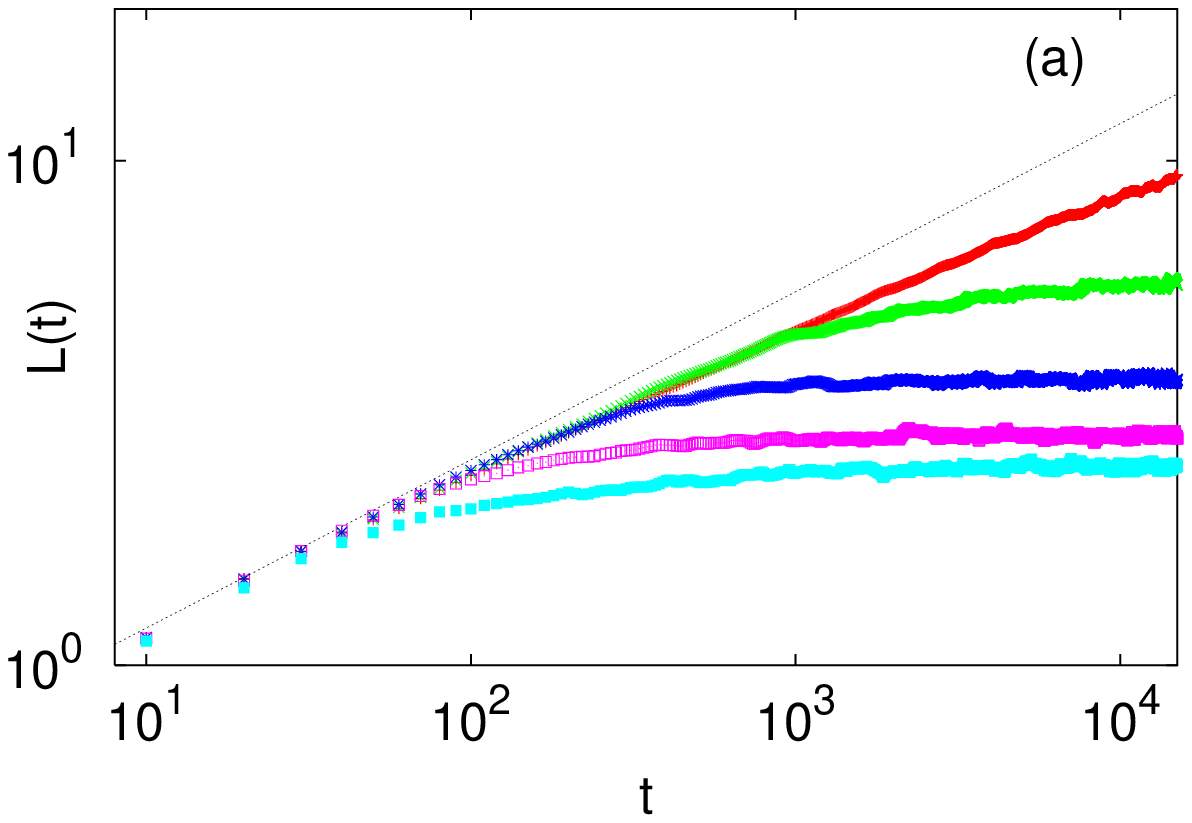}\hfill
\includegraphics[draft=false, scale=0.32,clip=true]{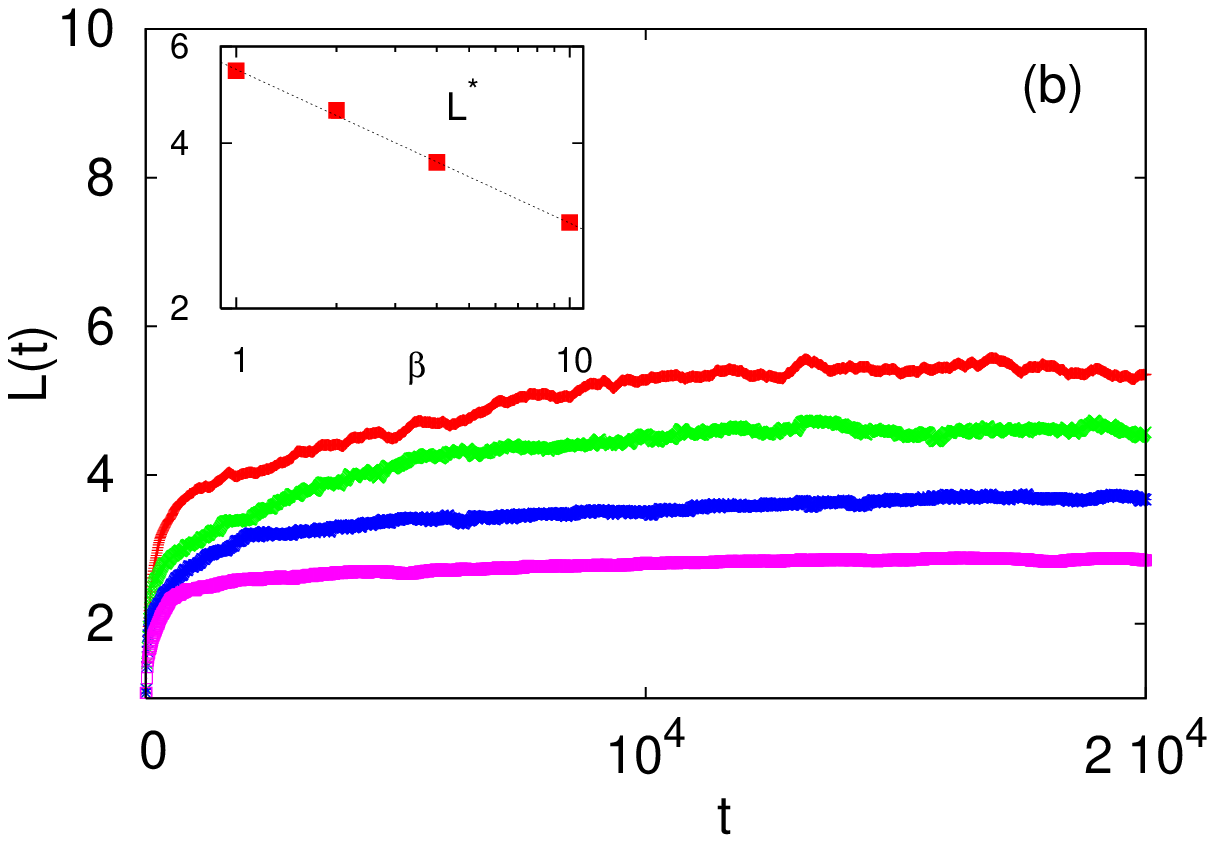}
\caption{(color online) Results of DNS in the passive case.  (a) $L$
vs $t$ with (from top) $\beta=0.25,0.5,1,2,4$. Forcing
parameters are $k_f=70$ and $F=3 \times10^{-5}$. The straight line
has slope $1/3$. (b) Same as (a), in linear scale, for
a frozen velocity field (see text), from top $\beta=1,2,4,10$. Inset:
$L^*$ vs $\beta$; the straight line corresponds to $L^* \sim
\beta^{-0.28}$, point size is of the order of the statistical error.
DNS were performed with hyper dissipation $-\nu_2\Delta^2\bm v$
with $\nu_2=10^{-7}$, and friction coefficient $\alpha=0.1$.  The
parameters of Eq.~(\ref{eq:ch}) are $\xi=0.015, \Gamma=0.02$.}
\label{fig:length_passive}
\end{figure}
 Previous studies stressed the importance of Lagrangian Chaos in the
coarsening arrest phenomenon~\cite{BBK01}. Now, in order to elucidate
this point, we discuss a non-chaotic example. It is well known (see,
e.g., \cite{CFPV91}) that two-dimensional stationary flows do not
generate chaotic trajectories.  We have thus integrated (\ref{eq:ch})
in a frozen configuration of the turbulent velocity field: ${\bm
v}({\bm x},t)={\bm v}({\bm x})$. As shown in
Fig.~\ref{fig:length_passive}b, domain growth is strongly weakened and
finally arrested, even in this non-chaotic flow.  For moderate
velocity intensities, $L(t)$ still grows in time, but with a much
slower scaling law than the dimensional prediction for fluids at rest,
$t^{1/3}$.  This slowing down is probably due to a different growth
mechanism: after an initial transient, a slow process of droplet
passage among close domains is indeed observed. However, for high
enough intensities a complete stabilization of the domain length is
realized.  This suggests that the main ingredient for coarsening
arrest is the presence of local shears that overwhelm the surface
tension driving force.  The dependence of $L^*$ on the shear rate
(here naturally defined as $\beta$) is shown in the inset of
Fig.~\ref{fig:length_passive}b.  We find a power law behavior with
exponent $-0.28$ very close to the one observed in the active case and
in shear flows \cite{HMM95}, while in the chaotic case no clear
scaling is observed.  \\\indent To further support the marginal role
of Lagrangian Chaos in coarsening arrest, we report the results
obtained in a stationary regular cellular flow ($v_x\!\! =\!U
\sin({\rm K}x)\cos({\rm K}y)\,,\; v_y \!=\! -U \cos({\rm K}x)\sin({\rm
K}y)$) where $U$ fixes the velocity amplitude and $K$ the
characteristic scale.  As shown in Fig.~\ref{fig:snaps_gollub} (left),
for large intensities the order parameter is frozen into a {\it random
chessboard} pattern with a finite length.  At lower intensities a
growth much slower than in the absence of the flow is still visible
(Fig.~\ref{fig:snaps_gollub} (right)) coming from a slow droplet
migration from one cell to another.  At large $U$'s the shear between
the counter-rotating vortexes overwhelms the demixing induced by the
thermodynamic forces, breaking the domains which freeze into the
cells.\\
\begin{figure}[t!]
\includegraphics[draft=false, scale=0.28, clip=true]{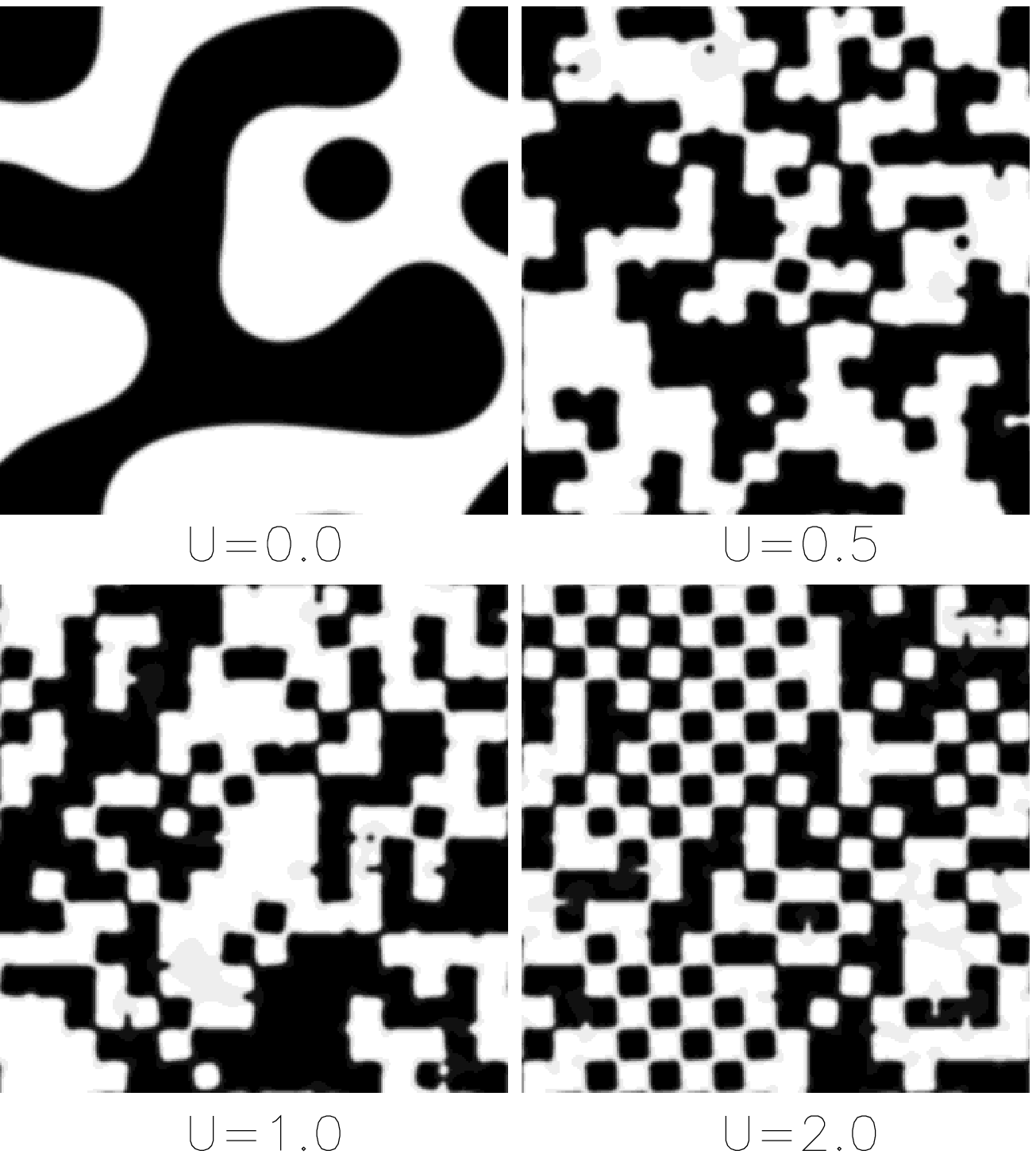} 
\hfill
\includegraphics[draft=false, scale=0.28, clip=true]{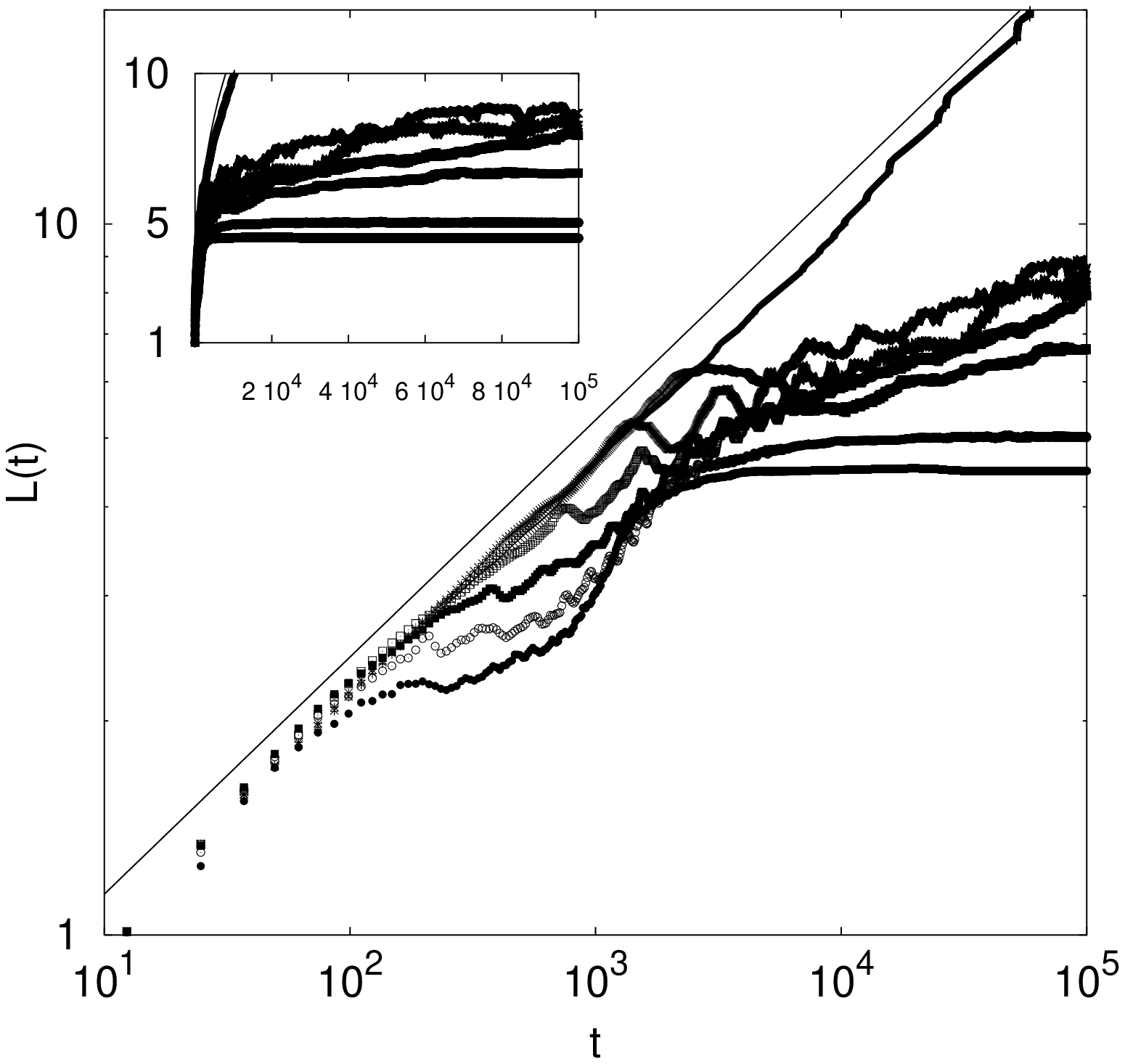} 
\caption{(left) Snapshots of $\theta$ at $t\!\!=\!6\!\times\!10^{4}$
 for the cellular flow with different $U$ and ${\rm K}=8$, here
 $\xi\!=\!0.018$, $\Gamma\!=\!0.1$. (right) $L$ vs $t$, from top to
 bottom $U\!=\!0, 0.125,0.25,0.5,1.0,2.0,4.0$.  Inset: the same in linear
 scale. }\label{fig:snaps_gollub}
\end{figure}
\indent In conclusion, we have shown that in the presence of an
external stirring coarsening process is slowed down both for active
and passive mixtures. We have also demonstrated that the phenomenon of
coarsening arrest, first predicted in \cite{RN81-AN84}, does not
necessarily require a chaotic flow, as suggested in \cite{BBK01}, but
is a consequence of the competition between thermodynamic forces and
stretching induced by local shears.  Our investigation in both active
and passive mixtures shows that this behavior is robust. Moreover we
found numerical evidence that the dependence of the arrest scale on
the shear rate follows a power law behavior with an exponent close to
the one measured in experiments and numerical simulations in pure
shear flows \cite{HMM95}. Our results might suggest the existence of a
mechanism independent of the nature of the flow in the coarsening
arrest. Further numerical and experimental investigations, with the
aim of clarifying the dependence of the arrest scale on the flow
properties, would be extremely interesting. \\

We are grateful to A.~Celani, G.~Gonnella and S.~Musacchio for useful
discussions.  We acknowledge partial support from MIUR Cofin2003
``Sistemi Complessi e Sistemi a Molti Corpi'', and EU under the
contract HPRN-CT-2002-00300.  MC acknowledges the MPIPKS for
computational resources.


\begin{thebibliography}{99}
\bibitem{nota2} Eq.~(\ref{eq:bilancio}) is obtained by averaging
  (\ref{eq:ch}) and (\ref{eq:ns}) after multiplication by $\mu$ and
  $\bm v$ respectively. The use of homogeneity, Gaussianity and
  $\delta$-correlation of the forcing gives the result.
\end{thebibliography}
\end{document}